\documentclass[11pt]{article}
\usepackage{graphicx}
\usepackage{cite}
\usepackage{amssymb,amsmath,amscd,amsthm}
\usepackage{amsfonts}
\usepackage{amsthm}
   \usepackage{float}
\usepackage{xcolor}
 \usepackage{pifont}

 \usepackage{hyperref}
\hypersetup{
   colorlinks   =  true,
    linkcolor    = blue,
    citecolor    = red,
     urlcolor	=blue,     
}

\newtheorem{proposition}{Proposition}

\newcommand\be{\begin{equation}}
\newcommand\ee{\end{equation}}

\newcommand{\PPhi}{{\Theta}} % calssical structure function

\newcommand{\cH}{\mathcal{H}} %Caligraphic H
\newcommand{\hcH}{ \hat{\mathcal{H}} } %hat  Caligraphic H
\newcommand{\hcC}{\hat{\mathcal{C}}} %hat Caligraphic C
\newcommand{\cC}{\mathcal{C}} %Caligraphic C
\newcommand{\hcK}{\hat {\mathcal{K}}} %hat Caligraphic K
\newcommand{\B}{{B}} % number operator B

\newcommand{\hB}{{\hat B}} %hat B
\newcommand{\hbm}{{\hat b}} %hat b
\newcommand{\hbp}{{\hat b}^\dagger} %hat b dagger

 \newcommand{\ri}{\imath} %  i complex
\def\>#1{{\mathbf #1}} %bold

\newcommand{\hI}{{\hat I}} %hat I
\newcommand{\hq}{{\hat q}} %hat q  
\newcommand{\hp}{{\hat p}} %hat p  

\newcommand{\n}{\mathrm{n}} %n
\newcommand{\fn}{\mathfrak{n}} % mathfrak principal quantum number

\newcommand{\mmu}{\mu} %  mu
  \newcommand{\nnu}{\nu} %  nu

\usepackage{color}

\begin{document}

\thispagestyle{empty}
\begin{center}
\Large{\textbf{Generalized classical and quantum Zernike Hamiltonians}}
\end{center}
\vskip 0.5cm

\begin{center}
\textsc{F.J.~Herranz$^{1,}\!\!\!~\footnote{
 Based on the contribution presented at ``The XIII International Symposium on {\em Quantum Theory and Symmetries (QTS-13)}",
28 July - 1 August, 2025, 
Yerevan State University,
 Yerevan, Armenia}$, A.~Blasco$^{1,\sharp}$, R.~Campoamor-Stursberg$^{2,\star}$, \\[3pt]
 I.~Gutierrez-Sagredo$^{3,\flat}$,   D.~Latini$^{4,\dagger}$ and I.~Marquette$^{5,\ddagger}$
 }
\end{center}

\begin{center}
 $^1$  {Departamento de F\'isica, Universidad de Burgos,
09001 Burgos, Spain}\\[4pt]
 $^2$  {Instituto de Matem\'atica Interdisciplinar \& Departamento de \'Algebra, Geo\-metr\'{\i}a \\ y Topolog\'{\i}a, Universidad Complutense de Madrid, 28040 Madrid,  Spain}\\[4pt] 
 $^3$ {Departamento de Matem\'aticas y Computaci\'on, Universidad de Burgos,\\ 
09001 Burgos, Spain}\\[4pt]
  $^4$   {Dipartimento di Matematica ``Federigo Enriques", Universit\`a degli Studi di Milano\\  \& INFN Sezione di Milano, 20133 Milano, Italy}\\[4pt]
   $^5$  {Department of Mathematical and Physical Sciences, La Trobe University, Bendigo,\\ VIC 3552, Australia}
\end{center}
 
    \smallskip

  { {
\qquad {\small $^*$\href{mailto:fjherranz@ubu.es}{\texttt{fjherranz@ubu.es}}\quad   $^\sharp$\href{mailto:ablasco@ubu.es}{\texttt{ablasco@ubu.es}}
\quad  $^\star$\href{mailto:rutwig@ucm.es}{\texttt{rutwig@ucm.es}}\quad  $^\flat$\href{mailto:igsagredo@ubu.es}{\texttt{igsagredo@ubu.es}}\\[2pt]   
\phantom{\qquad\qquad\qquad\qquad} $^\dagger$\href{mailto:danilo.latini@unimi.it}{\texttt{danilo.latini@unimi.it}}\quad   $^\ddagger$\href{mailto:i.marquette@latrobe.edu.au}{\texttt{i.marquette@latrobe.edu.au} } }}

\bigskip

\begin{abstract}
\noindent
 A superintegrable generalization of the classical and quantum Zernike systems is reviewed. The corresponding Hamiltonians are endowed with higher-order integrals and can be interpreted as higher-order superintegrable perturbations of the 2D  spherical (Higgs), hyperbolic, and  Euclidean harmonic oscillators. As a new result, the complete polynomial Higgs-type symmetry algebra   of  the generalized classical system is presented. For the generalized quantum system, the symmetry algebra and the spectra are  provided for a representative case.  
 \end{abstract}

\bigskip
\bigskip

integrable perturbations;  polynomial symmetry algebras; eigenvalue spectra

\newpage

%%%%%%%%%%%%%%%%%%%%%%%%%%%

\section{Introduction}

The quantum Zernike system \cite{Zernike} is defined by the  Hamiltonian given by~\cite{PSWY2017}
   \begin{equation}
\hcH_{\rm Zk} =  \hp_1^2 + \hp_2^2 -\ri \beta ( \hq_1 \hp_1+  \hq_2  \hp_2  )+\alpha ( \hq_1 \hp_1+  \hq_2  \hp_2  )^2\, ,
\label{a1}
\end{equation}
where $\hq_i$ and $\hp_i$ are quantum position and momentum operators with canonical  commutators  $ [\hq_i,\hp_j]= \ri \,\delta_{ij}$,  $\alpha$ and $\beta$ are two real parameters, and $\hbar=1$, thus providing the eigenvalue equation  $\hcH_{\rm Zk}  \psi(  {q_1,q_2} ) = E  \psi(  {q_1,q_2} )$. Beyond its relevance in quantum optics and Zernike
polynomials \cite{PSWY2017,Zpolynomials1,PWY2017a,Atakishiyev2019,Zpolynomials2},  the Zernike system turns out to be superintegrable, i.e., it admits two algebraically independent quantum observables that commute with the Hamiltonian $ \hcH_{\rm Zk}$, so that the spectrum exhibits maximal degeneracy.  
Moreover,  the case with $\alpha=-1$ and $\beta=-2$ is the only one that provides a self-adjoint  operator $ \hcH_{\rm Zk}$  under the inner product over the unit disk,  which was thoroughly studied   in \cite{PSWY2017}. Furthermore,   the classical counterpart $ \cH_{\rm Zk}$ of $ \hcH_{\rm Zk}$ (\ref{a1}), expressed  in terms of  canonical variables $q_i$ and $p_i$  with Poisson brackets $\{q_i,p_j\}=\delta_{ij}$, has also been  developed in  \cite{PWY2017c}. Clearly,   $\cH_{\rm Zk}$ determines a classical superintegrable system, with constants of motion  quadratic in the momenta, and therefore all bounded trajectories are   closed and  periodic.

The aim of this contribution is to review a superintegrable generalization of  the classical and quantum Zernike Hamiltonians,  entailing higher-order integrals, recently constructed in   \cite{BGSH2023} and \cite{Campoamor2025}, respectively. In addition, as a new result not included in  \cite{BGSH2023}, we present the complete polynomial symmetry algebra for the generalized classical system, whereas for the generalized quantum system, only a conjecture has been achieved so far.

%%%%%%%%%%%%%%%%%%%%%%%%%%%

\section{Generalized classical Zernike Hamiltonians}

Let us consider the  following  generalization of   $\cH_{\rm Zk}$ as   introduced in  \cite{BGSH2023}:
\begin{equation}
\label{a2}
\mathcal{H}_N=p_1^2+p_2^2+\sum_{n=1}^N \gamma_n (q_1 p_1+q_2 p_2)^n \, ,
\end{equation}
where  $N \in \mathbb{N}^\ast$ and $\gamma_n$ are arbitrary coefficients which can be real or imaginary   numbers. For $N = 2$,  $\gamma_1=-\ri \beta$, and $\gamma_2=\alpha$, $\mathcal{H}_2$ reduces to $\cH_{\rm Zk}$. The superintegrability property of $\mathcal{H}_N$ is characterized as follows.
\begin{proposition}
\label{prop1}\cite{BGSH2023}
For any $N$ and any values of the parameters $\gamma_n$, 
the Hamiltonian  $\mathcal{H}_N$ (\ref{a2}) Poisson-commutes with the angular momentum $\mathcal{C}= q_1 p_2 -q_2 p_1$ and the  constant of  motion  $ \mathcal{I}_N=\mathcal{I}_N(q_1,p_1,q_2,p_2)$  given by
\begin{equation}
\label{a3}
\mathcal{I}_N=p_2^2+\sum_{n=1}^N \gamma_n \sum_{j=0}^{\varphi(n)} p_2^{n-j}p_1^j Q^{(n-j,j)}(q_1, q_2) \, ,\quad  	\varphi (n) = 
	\begin{cases}
		n-2 & n\ \text{even} \\
		n-1 & n\ \text{odd} \\
	\end{cases}\, ,
\end{equation}
where $Q^{(n-j,j)}  (q_1,q_2)$ are homogeneous polynomials of degree $n$, whose explicit expressions can be found in \cite{BGSH2023}. The set $\{\mathcal{H}_N,\mathcal{C}, \mathcal{I}_N\}$ is formed by three functionally independent functions.
\end{proposition}
 Observe that starting from the free motion in the Euclidean plane $(q_1,q_2)\in \mathbb R^2\equiv \mathbf{E}^2 $,  the case with $N=1$ corresponds to add a linear momentum-dependent potential such that the constant of  the motion  $\mathcal{I}_1$ is quadratic, while the case $N=2$ leads to $\cH_{\rm Zk}$ with $\mathcal{I}_2$ being also quadratic.  By contrast, for $N\ge 3$, the $ \gamma_n$-terms provide higher-order potentials and $\mathcal{I}_N$ is of $N$th-order in the momenta. Recall that $\cH_2\equiv \cH_{\rm Zk}$  can alternatively be interpreted as an  oscillator on a space of constant curvature $\kappa$ with frequency $\omega$. Let us  set   $\gamma_1=  2\ri \omega$, $\gamma_2=-\kappa$, and apply the canonical transformation introduced in \cite{BGSH2023,Fordy2018}  from $(q_1,q_2,p_1,p_2)$ to the variables $ (\rho, \phi, p_\rho, p_\phi)$, where 
$\rho$  is the distance between the particle and the origin on the curved space  and $\phi\in[0,2\pi)$, finding that
 \be
 \label{a4}
\cH_{2} =   p_\rho^2 + \frac {\kappa 
\, p_\phi^2 } {  \sin^{2}(\sqrt{\kappa}  \rho)}   +\omega^2\,\frac{ \tan^2(\sqrt{\kappa}\rho)}{\kappa}\,  .
 \ee
 In this explicit form, the Hamiltonian $\cH_{2} $ covers the spherical (or Higgs~\cite{Higgs}) oscillator on $ \mathbf{S}^2$ $(\kappa>0)$, the hyperbolic one on $ \mathbf{H}^2$  $(\kappa<0)$ and the usual   harmonic oscillator on $ \mathbf{E}^2$  $( \kappa= - \gamma_2=0 )$. From this viewpoint, Proposition~\ref{prop1} determines superintegrable perturbations of such curved oscillators for $N\ge 3$.

To deduce the symmetry algebra of $\mathcal{H}_N$ (\ref{a2}), let us consider the  constant  of motion 
$\mathcal{I}'_N:=\mathcal{I}_N ( q_2,p_2,q_1,p_1)$ provided by   $\mathcal{I}_N$ (\ref{a3}) under the interchange $1\leftrightarrow 2$.   Thus, we have a   set of four constants $\{\mathcal{H}_N, \mathcal{I}_N, \mathcal{I}'_N, \mathcal{C}\} $ subjected  to the relation \cite{BGSH2023}
\begin{equation}
	\mathcal H_N=\mathcal I_N + \mathcal I_N' + \!\! \sum_{k=1}^{\varphi(N+1)/2} \!\!  (-1)^{k} \,  \gamma_{2k}\,\mathcal C^{2k}  \, .
 \label{a5}
\end{equation}
The complete symmetry algebra for $\mathcal{H}_N$ is established by a new proposition.
 \begin{proposition}
\label{prop2}
The constants of motion defined  by \begin{equation}
 \label{a6}
 \mathcal{L}_1:= \mathcal{C}/2\, , \quad  \mathcal{L}_2:=\bigl(\mathcal{I}'_N  -  \mathcal{I}_N \bigr)/2\, , \quad \mathcal{L}_3:= \{  \mathcal{L}_1,\mathcal{L}_2\} \, ,
\end{equation}
close  in  the  Poisson brackets  $\{ \mathcal{L}_1,\mathcal{L}_2\}=\mathcal{L}_3$,  $ \{ \mathcal{L}_1,\mathcal{L}_3\}=-\mathcal{L}_2$ and
\begin{gather}
  \{ \mathcal{L}_2, \mathcal{L}_3\}=- \sum_{n=1}^{N}  n\,  \PPhi_{N,n}(\mathcal{H}_N) \left(2\mathcal{L}_1\right)^{2n-1}  \, \quad \mbox{where} \label{a7} \\
 \PPhi_{N,n}(\mathcal{H}_N)   := \frac{ {\gamma^2_{ n}}}{2}-(-1)^{ n}\gamma_{2n}\mathcal{H}_N\delta_{1,\operatorname{sgn}(N-2n+1)}+  \!\! \!\sum_{s=1}^{\min\left(n-1,N-n\right)}\!\!(-1)^s \gamma_{ n-s}\gamma_{ n+s} \, ,\nonumber
\end{gather}
such that the Kronecker delta is $\delta=1$ whenever $N-2n+1\ge 1$.
\end{proposition}
  This   result confirms the conjecture  in  \cite[Eq.~(4.2)]{BGSH2023},   that for arbitrary $N$ the symmetry algebra of $\mathcal{H}_N$ (\ref{a2}) is a polynomial Higgs-type algebra of order $2N-1$. The   cubic Higgs algebra \cite{PWY2017c,Higgs} is recovered for the Zernike system  for $N = 2$.
 
 Propositions~\ref{prop1} and \ref{prop2} are illustrated in Table \ref{table1} by considering the case  $N=4$,
 where  $\mathcal I_4$ (\ref{a3}),  relation (\ref{a5}), structure functions $  \PPhi_{4,n}$ and Poisson bracket (\ref{a7}) are explicitly displayed.

  \newpage

 %%%%%%%%%%%%%%%%%%%%%%%%%%%

\begin{table}[htbp]
{
\caption[]{The generalized classical Zernike system for $N=4$}   
  \begin{center}
\begin{tabular}{@{}l@{}}
\hline
\hline
\\[-6pt]
 $\ \mathcal I_4=p_2^2+\gamma_1 Q^{(1,0)} p_2+  \gamma_2 Q^{(2,0)} p_2^2   + \gamma_3\left(Q^{(3,0)} p_2^3+Q^{(2,1)} p_2^2p_1+  Q^{(1,2)} p_2p_1^2 \right)        \phantom{\ }    $  \\[4pt]
   $\phantom{\  \mathcal I_4=p_2^2}  + \gamma_4\left(Q^{(4,0)} p_2^4+Q^{(3,1)} p_2^3p_1+  Q^{(2,2)} p_2^2p_1^2 \right)        $  \\[4pt]
   $\phantom{\ \mathcal I_4} =p_2^2+\gamma_1 q_2 p_2 +  \gamma_2 (q_1^2+q_2^2)p_2^2  + \gamma_3\bigl( q_2^3 p_2^3+(q_1^3+ 3 q_1 q_2^2) p_2^2p_1-q_2^3 p_2p_1^2 \bigr)       $  \\[4pt]
   $\phantom{\ \mathcal I_4=p_2^2}  + \gamma_4  \bigl( ( q_2^4-q_1^4) p_2^4+4 ( q_1^3 q_2+   q_1 q_2^3  ) p_2^3p_1+  (q_1^4-q_2^4) p_2^2p_1^2 \bigr)         $  \\[6pt]
   $\  \mathcal H_4=\mathcal I_4+ \mathcal I_4' - \gamma_{2}\,\mathcal C^{2}+\gamma_{4}\,\mathcal C^{4}    $  \\[6pt]
        $\   \PPhi_{4,1}=  \frac 12 \gamma_1^2 + \gamma_2  \mathcal H_4 $\qquad   $  \PPhi_{4,2}=  \frac 12 \gamma_2^2 -\gamma_1\gamma_3- \gamma_4  \mathcal H_4 $\\[6pt]
        $ \  \PPhi_{4,3}=  \frac 12 \gamma_3^2 - \gamma_2 \gamma_4 $  \qquad    $  \PPhi_{4,4}=  \frac 12 \gamma_4^2 $ \\[8pt]
    $ \  \{ \mathcal L_2, \mathcal L_3\}=-\big( \gamma_1^2 + 2 \gamma_2 \mathcal H_4\big)\mathcal L_1- \big(   \gamma_2^2- 2 \gamma_1\gamma_3- 2\gamma_4 \mathcal H_4\big)  (2\mathcal L_1)^3   $  \\[6pt]
     $\phantom{\  \{ \mathcal L_2, \mathcal L_3\}=}   - \tfrac 32 \big(  \gamma_3^2 -2\gamma_2\gamma_4   \big)(2 \mathcal L_1 )^5- 2\gamma_4^2(2 \mathcal L_1 )^7  $  \\[7pt]
\hline
\hline
\end{tabular}
  \end{center}
\label{table1}
}
\end{table}
  
   Note that, since the four parameters $\gamma_n$ are arbitrary, the case $N=3$ arises  as a byproduct for $\gamma_4=0$, the Zernike system with $N=2$    for  $\gamma_4=\gamma_3=0$, the harmonic oscillator with $N=1$  for $\gamma_4=\gamma_3=\gamma_2=0$  ($\gamma_1=  2\ri \omega$),  reducing to free (Euclidean) motion  when all $\gamma_n=0$.

   We recall that  a superintegrable extension of the Hamiltonian \eqref{a2} has been proposed in~\cite{Gonera2024} in terms of generic analytical functions $F (q_1 p_1 + q_2 p_2 )$, and that for $N=2$, the  issue of the imaginary $(\gamma_1=-\ri \beta)$-term in the Hamiltonian has been addressed in~\cite{Nerssesian} for both the classical and quantum systems.

%%%%%%%%%%%%%%%%%%%%%%%%%%%

\section{Generalized quantum Zernike Hamiltonians}

The generalization of the quantum Zernike system $\hcH_{\rm Zk}$ (\ref{a1}),  denoted  $\hcH_N$,  has been proposed in~\cite{Campoamor2025}   as the same formal function $\cH_N$ (\ref{a2}) written in terms of quantum position  and momentum operators  $(\hq_i,\hp_i)$. The Hamiltonian 
$\hcH_N$ commutes with the  quantum angular momentum operator 
$\hcC = \hq_1 \hp_2 - \hq_2 \hp_1$, so that an additional quantum integral $\hI_N$ must be obtained to prove superintegrability. Due to ordering problems this is by no means a trivial task and it has been solved in~\cite{Campoamor2025}  for $N\le 5$, obtaining two sets $\big\{\hcH_N,\hcC,\hI_N \big\}$ and $\big\{\hcH_N,\hcC,\hI'_N\big \}$  formed   by three algebraically independent operators. The integrals $\big\{\hcH_N,\hcC,\hI_N ,\hI'_N\big \}$, which satisfy an algebraic dependence relation,   allows us to deduce the  symmetry algebra  and spectrum by an algebraic procedure summarized in four steps:
\smallskip

\noindent
 (i)  From   $\big\{\hcC,\hI_N ,\hI'_N \big\}$ define
\be
 \hcK_1:=\hcC\, ,  \quad 
  \hcK_2:=  \big(\hI_N'-\hI_N\big)/2 \, , \quad
  \hcK_3:= \big[ \hcK_1, \hcK_2\big] \, ,
  \label{b1}
\ee
 which close on a polynomial deformation of $\mathfrak{sl}(2,\mathbb R)\simeq \mathfrak{so}(2,1)$.
\\

\noindent
 (ii) Introduce  number and ladder operators $\big\{ \hcK,\hcK_+,\hcK_-\big\}$ from $\big\{\hcK_1,\hcK_2,\hcK_3 \big\}$ via a (nonlinear) change of basis fulfilling the commutators
\be
\bigl[\hcK,\hcK_\pm\bigr]=\pm\hcK_\pm \, ,\quad \bigl[\hcK_-,\hcK_+\bigr]=\Phi\bigl(\hcH_N,\hcK+  {\rm I} \bigr)-\Phi\bigl(\hcH_N,\hcK  \bigr) \, ,
\label{b2}
\ee
where ${\rm I}$ is the identity operator and $\Phi$ is a structure operator which   factorizes as 
$$
\hcK_+\hcK_-\equiv \Phi\bigl(\hcH_N,\hcK\bigr)=\Phi_1\bigl(\hcH_N,\hcK\bigr)\Phi_2\bigl(\hcH_N,\hcK\bigr)\, ,
$$
determining   a $(2N-1)$th-order polynomial symmetry algebra of Higgs-type   $\mathfrak{sl}^{(2N-1)}(2,\mathbb R)$.

\noindent
 (iii)    A deformed oscillator algebra arises in a basis $\big\{\hB,\hbm,\hbp\big \}$ by setting
$$ \hB:= \hcK- u\,  {\rm I}\, , \quad \hbm:=\hcK_- \, , \quad
 \hbp:=\hcK_+  \, ,
 $$
  where  $u$ is a  constant. The commutation relations are  
   \be
 \bigl[\hB,\hbp\bigr]=\hbp \,, \quad \bigl[\hB,\hbm\bigr]=-\hbm \, , \quad \bigl[\hbm,\hbp\bigr]=\Phi\bigl(\hcH_N,\hB+(u+1) {\rm I}\bigr)-\Phi\bigl(\hcH_N,\hB+ u\,{\rm I}\bigr)\, .
 \label{b3}
 \ee
 (iv) And using the eigenvalues $E$ of $\hcH_N$ and $\B$ of $\hB$,  consider a  finite-dimensio\-nal   representation of (\ref{b3}) 
$$
\Phi=\Phi(\B,E,u)=\Phi_1(\B,E,u)\Phi_2(\B,E,u)
$$
subject to the conditions 
$$ \Phi(0,E,u)= 0\quad \mbox{and}\quad \Phi(\n+1,E,u)=0 \,  \quad
\mbox{for}\quad \n\ \in \{1,2,\dots\} \, .
 $$
The solutions $u=u(\n)$ and $E=E(\n)$ lead to $ \Phi(\B,E(\n),u(\n))\equiv \Phi(\B,\n)$ for $\B\in\{ 1,\dots,\n\}$ and $\n$  determines the principal quantum number $\fn$.
   
 We present the main results for $\hcH_4$ in Table~\ref{table2}: the   integral $\hI_4 $,  the relation among $\big\{\hcH_4,\hcC,\hI_4 ,\hI'_4 \big\}$, the generators $\big\{ \hcK,\hcK_+,\hcK_-\big\}$ from (\ref{b1}), the structure operator $\Phi=\Phi_1\Phi_2$ in (\ref{b2}), and the two types I and II of solutions   for the representation-dependent constant $u(\n)$ and the  spectrum $E=E(\n)$   by applying step (iv).

 %%%%%%%%%%%%%%%%%%%%%%%%%%%

\begin{table}[htbp]
\caption[]{The generalized quantum Zernike system for $N=4$}   
  \begin{center}
\begin{tabular}{@{}l@{}}
\hline
\hline
\\[-6pt]
 $\ \hI_4  =  \hp_2^2+ \gamma_1\hq_2\hp_2      +\gamma_2\bigl(   (\hq_1^2+\hq_2^2  ) \hp_2^2  -\hcC^2 \bigr) $  \\[4pt]
   $\phantom{\  \hI_4  =  \hp_2^2}  +\gamma_3\bigl(  \hq_2^3( \hp_2^3- \hp_1^2\hp_2)+( \hq_1^3+ 3 \hq_1\hq_2^2 ) \hp_1  \hp_2^2- 3 \ri \hq_2^2\hp_2^2- 3 \ri \hq_1 \hq_2\hp_1\hp_2-\hq_2  \hp_2\bigr)       $  \\[4pt]
       $\phantom{\  \hI_4  =  \hp_2^2}  +\gamma_4\bigl( (\hq_2^4-\hq_1^4)( \hp_2^4-\hp_1^2 \hp_2^2  ) +4( \hq_1^3\hq_2+  \hq_1\hq_2^3)\hp_1\hp_2^3 -6\ri (\hq_2^3+ \hq_1^2\hq_2) \hp_2^3     $  \\[4pt]
         $\phantom{\  \hI_4  =  \hp_2^2}\qquad\quad    -6\ri (\hq_1^3+\hq_1\hq_2^2 )\hp_1 \hp_2^2 -4(\hq_1^2+\hq_2^2)\hp_2^2  +4\hcC^2\bigr)  $  \\[6pt]
   $\  \hcH_4 =   \hI_4+  \hI_4' - 4 \gamma_4 \hcC^2 +\gamma_4 \hcC^4  $  \\[6pt]
  $\  \hcK := \frac{1}{2 }  \hcK_1 \qquad
\hcK_\pm :=   \hcK_2\pm\frac 12 \hcK_3- \big(  \frac{1}{2}\gamma_2-2 \gamma_4  \big) \hcK_1^2   $  \\[6pt]
          $\  \Phi_1(\hcH_4,\hcK)  =\frac 14 \bigl( \hcH_4 - 2\ri \gamma_1 \hcK +4 \gamma_2 \hcK^2+ 8\ri \gamma_3 \hcK^3- 16 \gamma_4 \hcK^4 \bigr)$\\[4pt]
            $\  \Phi_2(\hcH_4,\hcK)  =\hcH_4+ 2\ri \gamma_1\bigl(\hcK  - {\rm I} \bigr) + 4\gamma_2\bigl(\hcK  - {\rm I} \bigr)^2-
 8\ri \gamma_3\bigl(\hcK  - {\rm I} \bigr)^3- 16 \gamma_4\bigl(\hcK  - {\rm I} \bigr)^4\ $\\[6pt]
 $\mbox{\ \rm Types I \& II:}\quad   u_{\rm I} = u_{\rm II}= -\frac \n 2  \qquad 
(\gamma_1,\gamma_2,\gamma_3,\gamma_4)=(-\ri \beta,\alpha,\ri \mmu,- \nnu)    $\\[4pt]
 $\mbox{\ \rm Type I:}  \quad\ \ \, E_{\rm I}(\n)= -    \bigl(\ri \gamma_1 \n+\gamma_2   \n^2 - \ri \gamma_3 \n^3-   \gamma_4 \n^4 \bigr)$\\[4pt]
$\phantom{\ \mbox{\rm Type I:}  \quad\ \ \, E_{\rm I}(\n)}
=-    \bigl(\beta \n+\alpha   \n^2 +\mmu \n^3+\nnu \n^4\bigr)$\\[4pt]
$\mbox{\ \rm Type II:}  \quad E_{\rm II}(\n)=     \ri \gamma_1 (\n +2)-\gamma_2   (\n+2)^2  - \ri \gamma_3 (\n+2)^3 +\gamma_4 (\n+2)^4$\\[4pt]
$\phantom{\ \mbox{\rm Type II:}  \quad E_{\rm II}(\n)}
 =     \beta (\n +2)-\alpha   (\n+2)^2  +\mmu (\n+2)^3 -\nnu (\n+2)^4$\\[6pt]
\hline
\hline
\end{tabular}
  \end{center}
\label{table2}
\end{table}
  
%%%%%%%%%%%%%%%%%%%%%%%%%%%
 
Note that  the classical counterpart  $I_4$,   $I_4'$  of the quantum  integrals  $ \hI_4$, $  \hI_4'$  in Table~\ref{table2}
are related to  $\mathcal I_4$, $\mathcal I_4'$  in Table~\ref{table1} as $I_4=\mathcal I_4-\gamma_2 \cC^2 +4\gamma_4 \cC^2$ and $I_4'=\mathcal I_4'$. 

Observe that the spectra is real whenever odd coefficients $\gamma_n$ are imaginary numbers and even ones take  real values, thus  we have  introduced real coefficients  $\{\beta,\alpha,\mmu,\nnu\}$. Then, such spectra can be interpreted as third-order (with $\mmu$) and fourth-order (with $\nnu$) superintegrable perturbations of the curved oscillators (\ref{a4}) with curvature $\kappa=-\gamma_2=-\alpha$.  In addition,   if we set $\alpha=-1$ and $\beta=-2$ and interpret $\n$ 
 as the principal quantum number $\fn$, we find two possible superintegrable perturbations of the spectrum of  the Zernike system~$\hcH_{\rm Zk}$  \cite{PSWY2017}:    \be
E_{\rm I}=\fn(\fn+2) -\mmu \fn^3 -\nnu \fn^4,\,\quad E_{\rm II}=\fn(\fn+2) +\mmu(\fn+2)^3-\nnu (\fn+2)^4\, .
     \ee
  Finally, we  remark that the spectrum for $\hcH_N$ should be deduced by solving the Schr\"odinger equation, at least for low values  $N=3,4$, which would show strong differences for the spherical and hyperbolic cases (see~\cite{Kuruannals} for $N=2$).

 %%%%%%%%%%%%%%%%%%%%%%%%%%%%%%%%%%%%%%%%%%%%
 
 \section*{Acknowledgements} 
 \small
  F.J.H.,  A.B., R.C.-S.~and I.G.-S.~were   supported by Agencia Estatal de Investigaci\'on (Spain) under    grant PID2023-148373NB-I00 funded by MCIN/AEI/ 10.13039/501100011033/FEDER EU,  and      Q-CAYLE Project,     Junta de Castilla y Le\'on and MCIN  (Spain),  with EU funds NextGenerationEU (PRTR C17.I1).  The research of D.L.~was supported by MUR -- Dipartimento di Eccellenza 2023--2027 (CUP G43C22004580005, project code DECC23\_012\_DIP) and by INFN--CSN4 (MMNLP project). D.L. is a member of GNFM, INdAM. I.M.~was supported by Australian Research Council Future Fellowship FT180100099.

 %%%%%%%%%%%%%%%%%%%%%%%%%%%%%%%%%%%%%%%%%%%%

\end{document}